\documentclass[11pt,aps,
superscriptaddress,prc,eprint,nofootinbib]{revtex4-2}
\pdfoutput=1
\usepackage{graphicx}
\usepackage{dcolumn}
\usepackage{bm}

\usepackage{verbatim}
\usepackage{empheq}

\usepackage{amssymb}

\usepackage{mathtools}
\usepackage{subfigure}
\usepackage{bbold}
\usepackage{yfonts}
\usepackage[colorlinks=true,linktocpage=true,linkcolor=blue,citecolor=blue]{hyperref}
\usepackage{placeins}
\usepackage{bm} 
\usepackage{nicefrac}
\usepackage{slashed}
\usepackage{marginnote}

\usepackage{amsmath,amsfonts,amsthm,amssymb}
\usepackage{graphics}
\usepackage{cancel}
\usepackage{multirow}
\usepackage{longtable}
\usepackage{color}
\usepackage{xcolor}
\usepackage[normalem]{ulem}
\usepackage{dsfont}

\newcounter{rownum} 
\usepackage{amssymb}

\usepackage{graphicx}
\usepackage{mathtools}
\usepackage{subfigure}
\usepackage{bbold}
\usepackage{yfonts}
\usepackage[colorlinks=true,linktocpage=true,linkcolor=blue,citecolor=blue]{hyperref}
\usepackage{placeins}
\usepackage{bm} 
\usepackage{nicefrac}
\usepackage{slashed}
\usepackage{marginnote}

\begin{document}
	
\title{Charm Quark Kinetics in Heavy Ion Collisions}

\author{Valeriya Mykhaylova}
\email{valeriya.mykhaylova@uj.edu.pl}
\affiliation{Institute of Theoretical Physics, University of Wrocław, plac Maksa Borna 9, PL-50204 Wrocław, Poland}
\affiliation{Institute of Theoretical Physics, Jagiellonian University, ul. prof. St. \L ojasiewicza 11, 30-348 Kraków, Poland}

\author{Krzysztof Redlich}
\email{krzysztof.redlich@uwr.edu.pl}
\affiliation{Institute of Theoretical Physics, University of Wrocław, plac Maksa Borna 9, PL-50204 Wrocław, Poland}
\affiliation{Polish Academy of Sciences PAN, Podwale 75, PL-50449 Wrocław, Poland}

\author{Chihiro Sasaki}
\email{chihiro.sasaki@uwr.edu.pl}
\affiliation{Institute of Theoretical Physics, University of Wrocław, plac Maksa Borna 9, PL-50204 Wrocław, Poland}
\affiliation{International Institute for Sustainability with Knotted Chiral Meta Matter (WPI-SKCM2), Hiroshima University, 1-3-1 Kagamiyama, 739-8526, Higashi-Hiroshima, Hiroshima, Japan}

\date{\today}

\begin{abstract}

We study the evolution of charm $(c)$ quarks in hot QCD matter with $N_f=2+1(+1)$ quark flavors by analyzing the charm production rate and the time dependence of their abundance. Microscopically, the system is described within a quasiparticle model, in which interactions among dynamical quarks and gluons are encoded in their effective masses with the running coupling constrained by lattice QCD data. We investigate $c$--quark kinetics in a longitudinally propagating perfect fluid as well as in a viscous medium undergoing (2+1)D expansion, and find that the charm production rate decreases monotonically across all medium formulations. In the $N_f=2+1+1$ scenario, charm production is systematically suppressed due to the effective mass of heavy quasiparticles. Assuming an initial charm yield given by the Statistical Hadronization Model, we solve the rate equation and compute the total charm abundance in hot QCD medium. For all descriptions considered, the charm quark number remains approximately conserved, consistent with existing experimental evidence.
\end{abstract}

\maketitle

\bigskip

\section{Introduction}

The behavior of heavy quarks in hot QCD matter provides a sensitive probe of the microscopic structure and dynamical properties of the quark–gluon plasma (QGP)~\cite{Gross:2022hyw}. Owing to their large masses, charm quarks are predominantly produced in initial hard scatterings and are not expected to reach full chemical equilibrium during the finite lifetime of the medium created in relativistic heavy-ion collisions (HIC)~\cite{Moore:2004tg}.~Nevertheless, their subsequent interactions with the QGP constituents encode valuable information about in-medium QCD dynamics, including the role of non-perturbative effects near the crossover region~\cite{Rajagopal:2025ukd}.

So far, the main focus of heavy-flavor physics has been on understanding various observables of charmed hadrons~\cite{Braun-Munzinger:2024ybd,Berwein:2024ztx,Dong:2019byy,Apolinario:2022vzg} and quarkonia~\cite{Rapp:2008tf,Singh:2025dfx,Zhao:2011cv,Ding:2012sp,Andronic:2025jbp}, as well as the transport of heavy quarks~\cite{Zhao:2023nrz,Yao:2020xzw,Beraudo:2023nlq}. These studies have primarily focused on energy loss~\cite{Peng:2024zvf,Dang:2023tmb,Xing:2023ciw,ALICE:2023jgm,Beraudo:2025nvq}, spatial diffusion~\cite{HotQCD:2025fbd,He:2022ywp,Altenkort:2023oms,Capellino:2023cxe,Altenkort:2023eav,Sambataro:2025obe,Scardina:2017ipo}, and collective flow~\cite{Zhao:2023ucp,CMS:2018loe}.

Comparatively less effort has been devoted to investigating the chemical kinetics of open heavy flavors in the expanding QGP~\cite{Song:2024hvv,Zhang:2007yoa,Levai:1997bi,Mykhaylova:2025mht}. In particular, the question of whether charm quarks can be thermally produced or annihilated in significant numbers during the deconfined phase remains an open issue. Addressing this problem requires a consistent treatment of both the microscopic production mechanisms and the macroscopic space--time evolution of the medium~\cite{Krishna:2025bll}. 
    
Early studies of charm production in the QGP suggested that secondary charm–anticharm pair creation is negligible at experimentally accessible energies, leading to an approximate conservation of the total charm yield throughout the evolution of the fireball~\cite{Graf:2018lok,Uphoff:2010sh}. However, the advances in lattice QCD (lQCD) calculations~\cite{Borsanyi:2016ksw,HotQCD:2025fbd}, effective kinetic approaches~\cite{Song:2024hvv}, and hydrodynamic modeling~\cite{Beraudo:2025nvq,Capellino:2023cxe} have renewed interest in revisiting this assumption. In particular, the increasing precision of experimental measurements of open-charm production at the Large Hadron Collider (LHC)~\cite{Bierlich:2023ewv,Braun-Munzinger:2024ybd,LHCb:2023cwu,LHCb:2022dmh} motivates a quantitative reassessment of charm quark kinetics, including possible deviations from chemical equilibrium.

An additional theoretical challenge arises from the treatment of charm quarks within the QGP.~In many phenomenological studies, charm quarks are treated as heavy impurities propagating through a thermalized background of light quarks and gluons\mbox{~\cite{Rapp:2008qc,Rapp:2009my,Scardina:2017ipo,Song:2024hvv}}. However, at sufficiently high temperatures, $c$ quarks may contribute non-negligibly to the thermodynamic properties of the medium itself~\cite{Borsanyi:2016ksw}. This raises the question of how the inclusion of dynamical charm degrees of freedom affects both the equation of state (EoS) and secondary charm-quark production.
    
In this work, we investigate the kinetics of charm quarks in hot QCD matter by analyzing the charm production rate and the time evolution of the total charm abundance. The medium is described within the quasiparticle model (QPM)~\cite{Bluhm:2004xn,Bluhm:2006yh,Bluhm:2007nu,Mykhaylova:2019wci,Mykhaylova:2020pfk}, in which interactions among quarks and gluons are encoded in their temperature-dependent effective masses governed by a running coupling constrained by lQCD data. This framework allows for a unified description of the QGP across a wide temperature range, from the perturbative regime at high temperatures down to the non-perturbative crossover region.

We consider two distinct descriptions of charm quarks in the QGP. In the first case, corresponding to QCD with $N_f=2+1$ quark flavors, $c$ quarks are treated as impurities with fixed masses, while the medium consists of thermalized quasiparticles (light and strange quarks, and gluons). In the second case, we extend the description to $N_f=2+1+1$ flavors, where charm quarks contribute to the EoS and therefore are fully incorporated as dynamical quasiquarks dressed by temperature-dependent effective masses. This setup enables a systematic assessment of the impact of thermalized charm degrees of freedom on charm quark production rate and time evolution of their abundance.

The space--time evolution of the QGP is modeled using both a simplified boost-invariant longitudinal propagation of perfect fluid (Bjorken flow), and a more realistic hydrodynamic expansion of viscous fluid in (2+1) dimensions. Starting from an initial charm yield determined by the Statistical Hadronization Model (SHM), we solve the rate equation governing the charm quark production and annihilation throughout the QGP lifetime. Further, we calculate the time evolution of the total number of charm quarks (the charm quark abundance) and assess its sensitivity to the medium composition and expansion dynamics.

The paper is organized as follows. In Sec.~\ref{sec:QPM}, we introduce the concept of the QPM and discuss how the effective running coupling and the quasiparticle masses are determined. Section~\ref{sec:evolution} describes the space--time evolution of the QGP, which includes modeling its temperature and volume  profiles. In Sec.~\ref{sec:rate} we present the formulation of the differential rate equation, focusing on the relevant cross sections involving $c$ quarks (Sec.~\ref{sec:cross}) and computing the total charm quark production rate (Sec.~\ref{sec:Rate-exp}).
The resulting time evolution of the charm abundance is discussed in Sec.~\ref{sec:yield}. Finally, in Sec.~\ref{sec:summary}, we summarize our findings and outline possible directions for future studies.

\section{Quasiparticle Approach \label{sec:QPM}}
	
To investigate the evolution of heavy quarks in a thermal bath of hot QCD matter with \mbox{$N_f=2+1(+1)$} quark flavors, we employ the QPM~\cite{Mykhaylova:2019wci,Mykhaylova:2020pfk,Bluhm:2004xn,Bluhm:2006yh,Bluhm:2007nu}, which  describes the system in terms of effective quasiparticle degrees of freedom (d.o.f.).~Microscopic interactions among the QGP constituents are encoded in dynamically generated quasiparticle masses through a temperature-dependent running coupling $G(T)$. The effective coupling is determined by requiring that the entropy density of the medium reproduces the  entropy density scaled by the temperature, $s/T^3$, obtained from lQCD simulations~\cite{Borsanyi:2013bia,Borsanyi:2016ksw}.

Within this effective kinetic framework, the hot QCD matter is described as an ideal gas of massive quasiparticles, thus its scaled entropy density is given by~\cite{Mykhaylova:2019wci}

\begin{equation}
	\label{e:entr0}
	\frac{s}{T^3} = \frac{1}{T^3}\sum_{i=g,l,s,(c)} s_i \, =\,  \frac{1}{T^3}\sum_{i=g,l,s,(c)} \frac{d_i}{2\pi^2} \int_0^\infty
	\!\!\! p^2 dp \frac{\left( \frac{4}{3}p^2{+}m_i^2 \right)}{E_i\, T}
	f_i^{\rm eq} \,,
\end{equation}
where the index $i$ denotes gluons $(g)$, light  quarks $(l)$, strange  quarks $(s)$, and, optionally, charm quarks $(c)$ in the $N_f=2+1+1$ case. Since the analysis is performed at vanishing quark chemical potential, $\mu = 0$, particles and antiparticles contribute equally. Consequently, the quark terms in Eq.~\eqref{e:entr0} must be counted twice. The spin--color degeneracy factors $d_i$ are given by $d_l=12$ for the two light quark flavors, $d_{s,c}=6$ for the strange  and charm quarks, and $d_g=16$ for gluons.

The phase-space distributions of the quasiparticles in thermal equilibrium are given by the standard Fermi--Dirac or Bose--Einstein distribution functions,
\begin{equation}
 \label{eq:distr}
	f_i^{\rm eq}= (e^{E_i/T}+S_i)^{-1} 
\end{equation}
where $S_{l,s,c}=1$ for fermions and $S_g=-1$ for bosons.  We assume that the quasiparticles propagate on the mass shell. Accordingly, the dispersion relation reads \mbox{$E_i=\sqrt{p^2+m_i^2}$}, with the momentum $p$ and the effective mass  of the quasiparticle $m_i\equiv m_i(T)$ defined as
\begin{gather}
	m_i^2(T)=(m_{i}^0)^2 +\Pi_i(T). \label{eq:mass}
\end{gather}
Here, $m_{i}^0$ denotes the current (bare) particle mass, for which we adopt \mbox{$m_g=0$ MeV}, \mbox{$m^0_l=5$ MeV}, \mbox{$m^0_s=95$ MeV}, and, for the $N_f=2+1+1$ case, $m^0_c=1.3$~GeV. In the QGP with  $N_f=2+1$, charm quarks are treated as impurities with a constant mass $M_c=1.3$~GeV, denoted by a capital letter to distinguish it from the temperature-dependent charm quark mass, $m_c(T)$, employed in the 2+1+1 scenario.

For~the self-energy $\Pi_i$, we employ the gauge-independent expressions computed within the hard thermal loop approach~\cite{Bluhm:2006yh,Pisarski:1989wb},
\begin{align}
	\label{e:piq}
	\Pi_{l,s,(c)}(T)  & =  2 \left(m_{l,s,(c)}^0 \sqrt{\frac{G(T)^2}{6}T^2}+
	\frac{G(T)^2}{6}T^2\right),\\
	\Pi_g(T) & = \left(3+\frac{N_f}{2}\right)\frac{G(T)^2}{6}T^2, \label{eq:pig}
\end{align}
where the number of quark flavors is set to $N_f=3$ or $4$, depending on the considered QGP structure. As discussed above, the effective running coupling $G(T)$ is determined numerically from the lQCD data~\cite{Borsanyi:2013bia,Borsanyi:2016ksw}, by simultaneously solving Eqs.~\eqref{e:entr0}--\eqref{eq:pig}.

It is worth emphasizing that previous applications of the QPM have demonstrated that the effective running coupling $G(T)$ and dynamical quasiparticle masses $m_i(T)$ yield a consistent description of hot QCD matter across a broad temperature range. In particular, quantities evaluated within the QPM show good qualitative and quantitative agreement with results obtained from various theoretical frameworks in their respective domains of validity. At high temperatures, the QPM reproduces perturbative QCD (pQCD) expectations for several transport coefficients, including the shear and bulk viscosities, while in the crossover region it retains essential nonperturbative QCD features and agrees with the alternative appropriate formulations, such as those based on the gauge-gravity duality~\cite{Mykhaylova:2019wci,Mykhaylova:2020pfk,Mykhaylova:2021cep}.

\begin{figure}[t]
	\centering
	\includegraphics[width=0.48\linewidth]{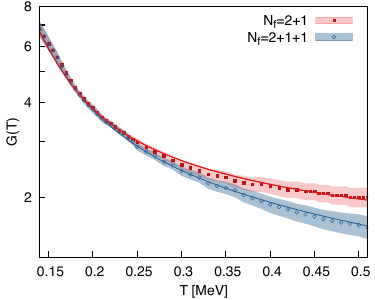}
	\includegraphics[width=0.48\linewidth]{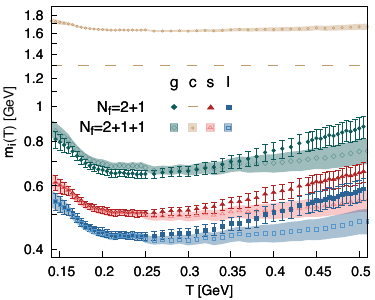}
	\caption{Left: Effective running coupling $G(T)$ as a function of temperature $T$ for hot QCD matter with $N_f=2+1$ (squares) and $N_f=2+1+1$ (circles). The shaded bands indicate the uncertainties propagated form the underlying lQCD data~\cite{Borsanyi:2013bia,Borsanyi:2016ksw}. Analytical fits to the corresponding data points are shown by solid lines. Right: Dynamically generated quasiparticle masses $m_i(T)$ as functions of temperature for $N_f=2+1$ (filled symbols) and $N_f=2+1+1$ (open symbols with shaded bands). Gluons $(g)$ are shown by diamonds, charm quarks ($c$)  by circles, strange quarks ($s$) by triangles, and light quarks $(l)$ by squares. The dashed line indicates the value of the bare charm quark mass, $M_c=1.3$ GeV.}
	\label{fig:GT}
\end{figure}

The left panel of Fig.~\ref{fig:GT} displays the effective running couplings $G(T)$ obtained within the QPM for systems with different quark contents. Both couplings exhibit similar temperature dependence over the studied temperature range and coincide at low $T$, reflecting the behavior of the underlying lQCD data~\cite{Borsanyi:2016ksw}. Charm quarks begin to contribute appreciably to the EoS above $T\simeq 300$~MeV. Consequently, at high $T$,  the system with thermalized charm quarks $(N_f=2+1+1)$ is characterized by a larger entropy density and a correspondingly weaker effective coupling compared to the \mbox{$N_f= 2+1$} case. A similar trend was reported in~\cite{Sambataro:2024mkr}, where the effective coupling 
$G(T)$ was extracted by fitting the energy density $\epsilon(T)$ of the same EoS used in the present study~\cite{Borsanyi:2016ksw}.

The right panel of Fig.~\ref{fig:GT} shows the temperature dependence of the dynamically generated quasiparticle masses $m_i(T)$ for $N_f=2+1$ and $N_f=2+1+1$ quark flavors.  All effective masses exhibit a shallow minimum slightly above $T_{\rm c}$, followed by an approximately linear increase with temperature. At low $T$, the effective masses of light and strange quasiparticles overlap, respectively, in both flavor contents, reflecting the behavior of the effective coupling. In contrast, a small shift in the quasigluon masses arises from the explicit flavor dependence of $m_g(T)$, see Eq.~\eqref{eq:pig}.~At higher~$T$, the weaker coupling in the $N_f=2+1+1$  system leads to reduced effective masses of the quasiparticles compared to the $N_f=2+1$ case. Charm quarks represent a special case, as their current mass dominates the self-energy contribution, resulting in a weak temperature dependence and an approximately constant shift of $m_c(T)$ relative to the constant charm mass $M_c$ used in the $N_f=2+1$ scenario.

The effective masses obtained  within the underlying QPM for $N_f=2+1+1$ are also consistent with the results from an alternative quasiparticle framework, the $QPMp$~\cite{Sambataro:2024mkr}. In this approach, the quasiparticle masses acquire an explicit momentum dependence, allowing them to  converge to the bare quark mass at high momentum. The $p$-dependence also increases the effective masses compared to those shown in Fig.~\ref{fig:GT} (right), enabling quantitative agreement with lQCD results for quark susceptibilities~\cite{Sambataro:2024mkr}.

\section{Space--Time Evolution of the QGP} \label{sec:evolution}
\subsection{Temperature \label{subsec:T}}
Before solving the rate equation (see Sec.~\ref{sec:rate}), we must specify the space--time evolution of hot QCD matter, since the rate equation involves explicit time derivatives.\\[-1.5em]

For the QGP with $N_f=2+1$ quark flavors, we consider two distinct evolution scenarios:

\textbf{(a)} As a first approximation, the system is treated as a boost-invariant perfect fluid undergoing purely longitudinal, one-dimensional  (1D) expansion. Neglecting dissipative effects, the system follows Bjorken flow~\cite{Bjorken:1982qr}, with the temperature evolving according to the Bjorken scaling solution,
\begin{eqnarray}
	T(\tau)= T_0 \Big(\frac{\tau_0}{\tau}\Big)^{1/3} \label{eq:Ttauideal},
\end{eqnarray}
where $\tau$ is the proper time and $T_0 \equiv T(\tau_0)$ is the initial temperature at the onset of the QGP evolution, which begins at the thermalization time $\tau_0$. For the initial condition we take \mbox{$T_0=0.624$ GeV}, \mbox{$\tau_0=0.2$ fm.}

\begin{figure}[t!]
	\centering
	\includegraphics[width=0.6\linewidth]{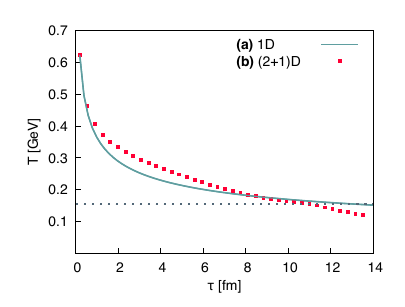}
	\caption{Temperature $T$ as a function of proper time $\tau$ for $\textbf{(a)}$ a longitudinally  propagating  perfect fluid (solid line) and $\textbf{(b)}$ a viscous medium expanding in (2+1) dimensions (squares). Both scenarios start from identical initial conditions, \mbox{$T_0=0.624$ GeV} and  \mbox{$\tau_0=0.2$ fm}. The dashed line indicates the pseudocritical temperature, \mbox{$T_{\rm c}=155$ MeV}.} 
	\label{fig:T_tau}
\end{figure}

\textbf{(b)} We also consider a more realistic scenario in which the system is described as a boost-invariant viscous fluid expanding in (2+1) dimensions, including transverse dynamics. For the temperature evolution, we assume an azimuthally symmetric profile and adopt the values at the center of the fireball obtained by~\cite{Auvinen:2020mpc}. The  QGP hydrodynamization begins at the same initial condition as in scenario $\textbf{(a)}$, namely \mbox{$T_0=0.624$ GeV,} \mbox{$\tau_0=0.2$ fm.}

The $T(\tau)$ profile used in scenario $\textbf{(b)}$ is obtained from a Bayesian analysis of relativistic HIC data~\cite{Auvinen:2020mpc}. The properties of hot QCD matter in that study are also consistent with those assumed in the present quasiparticle framework. In particular, the temperature-dependent specific shear viscosity, $(\eta/s) (T)$, obtained within the QPM~\cite{Mykhaylova:2019wci} is quantitatively consistent with the specific shear viscosity determined from hydrodynamic simulations (see Fig.~5 of Ref.~\cite{Auvinen:2020mpc}).~Moreover, both approaches rely on the same lQCD entropy density data for $N_f=2+1$~\cite{Borsanyi:2013bia}.

For QCD with $N_f=2+1+1$, the medium expansion is described using scenario $\textbf{(a)}$. This choice allows us to isolate and highlight the effects of treating charm quarks as dynamically dressed quasiparticles, in contrast to the $N_f=2+1$ QGP, where $c$ quarks are included as heavy impurities with a fixed mass. The corresponding analysis is presented in Sec.~\ref{sec:yield}.

Figure~\ref{fig:T_tau} shows the time evolution of the temperature for the two expansion scenarios described above.  Despite the different physical assumptions, namely, perfect versus viscous fluid and 1D versus (2+1)D expansion, the use of identical initial condition, $T_0(\tau_0)$,  results in good qualitative and quantitative agreement between the corresponding  $T(\tau)$ profiles.

A notable difference emerges in the crossover region. In scenario $\textbf{(a)}$, corresponding to a perfect Bjorken flow, the system reaches the pseudocritical temperature \mbox{$T_{\rm c}=155$~MeV} at \mbox{$\tau_{\rm fin} \simeq13$ fm}, whereas in $\textbf{(b)}$, where the QGP is described as a viscous medium expanding in (2+1)D, hadronization occurs earlier, at $\tau_{\rm fin} \simeq11$ fm. This behavior naturally follows from the inclusion of transverse expansion and viscous effects in  scenario \textbf{(b)}, which increase the effective volume and introduce energy dissipation, thereby accelerating the system's cooling.

The temperature range considered in this study spans the lifetime of hot QCD matter in thermal equilibrium,  from the thermalization temperature $T_0(\tau_0)$ down to the pseudocritical temperature $T_{\rm c}(\tau_{\rm fin})$, at which hadronization occurs. Therefore, temperatures below $T_{\rm c}$, or equivalently times beyond $\tau_{\rm fin}$ in each scenario, are shown for illustrative purposes only.

\subsection{Volume \label{subsec:V}}

To study the evolution of the charm abundance, i.e., the total number of charm quarks in the QGP, we must  specify both the temperature profiles $T(\tau)$ and the proper-time dependence of the system volume.

Following~\mbox{\cite{Zhang:2007yoa,Ko:2010zzc,Kumar:2014kfa,Sheikh:2021opp}},  the QGP volume per unit rapidity is modeled as
\begin{align}
	V(\tau)=\pi R^2 \tau, \label{eq:V}
\end{align}
where $R$ is the transverse radius of the thermalized fireball. Corresponding to the evolution scenarios introduced in Sec.~\ref{subsec:T}, we consider:

$\textbf{(a)}$ A perfect QGP undergoing purely longitudinal expansion (Bjorken flow) with a constant transverse radius, $R=7$~fm, resulting in a longitudinally stretching cylindrical volume.

$\textbf{(b)}$ Including transverse expansion, the radius of the viscous medium expanding in (2+1)D becomes time dependent and is parametrized as~\cite{Zhang:2007yoa,Ko:2010zzc}
\begin{gather}
	R(\tau)=R_0+\frac{a}{2}(\tau-\tau_0)^2, \label{eq:R_tau}
\end{gather}
where $R_0\equiv R(\tau_0)=7$~fm is the initial radius, and $a = 0.085\ \text{fm}^{-1}$ is the transverse acceleration. This choice of the parameters ensures that the volume given by Eq.~(\ref{eq:V}) reaches \mbox{$V_{\rm fin}=4997\pm455\ \text{fm}^3$} at  $\tau_{\rm fin}\simeq 11$ fm, corresponding  to the end of the QGP evolution. This matches the fireball volume per unit rapidity at midrapidity at the chemical freezeout ($T_{\text{cf}}=156.5$~MeV) used in the SHM for the most central $(0-10\%)$ Pb-Pb collisions \mbox{at $\sqrt{s_{NN}}=5.02$~TeV~\cite{Andronic:2021erx}.}

\begin{figure}[htb!]
	\centering
	\includegraphics[width=0.6\linewidth]{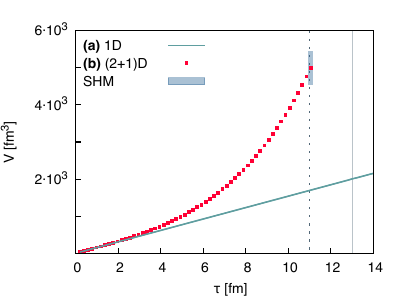}
	\caption{Volume of the hot QCD medium as a function of a proper time, $V(\tau)$, for different expansion scenarios: $\textbf{(a)}$ purely longitudinal propagation of a perfect fluid (solid line), $\textbf{(b)}$ expansion of a viscous medium in (2+1) dimensions (squares). Both evolutions start at  $\tau_0=0.2$~fm with an initial volume per unit rapidity $V_0\simeq 30.8\ \text{fm}^3$, determined from Eqs.~\eqref{eq:V}--\eqref{eq:R_tau}. The shaded band corresponds to the volume of the fireball per unit rapidity at chemical freezeout, \mbox{$V_{\rm fin}=4997\pm455\ \text{fm}^3$}, used in the SHM~\cite{Andronic:2021erx} (see main text for details).~The vertical lines indicate the times at which the QGP reaches the pseudocritical temperature, $T_{\rm c}=155$ MeV: solid line for scenario~$\textbf{(a)}$ and  dashed line for $\textbf{(b)}$. }
	\label{fig:V_tau}
\end{figure}

Figure~\ref{fig:V_tau} shows the time evolution of the QGP volume for the two expansion scenarios. With the identical initial condition, the  volumes coincide at early times.~Around \mbox{$\tau\simeq 3.5$ fm}, the difference between the two evolution schemes becomes apparent, leading to a final volume roughly three times smaller for an ideal Bjorken flow $\textbf{(a)}$ compared to a viscous fireball expanding in (2+1)D~$\textbf{(b)}$. As discussed above, scenario $\textbf{(b)}$ provides a more realistic description by incorporating both dissipative effects and transverse expansion, while also matching the final volume to values relevant for certain LHC collisions. The vertical lines indicate the endpoints of the QGP evolution at the pseudocritical temperature, corresponding to \mbox{$\tau_{\text{fin}}\simeq13$ fm} for $\textbf{(a)}$  and  $\tau_{\text{fin}}\simeq11$ for $\textbf{(b)}$, consistent with the temperature evolution shown in Fig.~\ref{fig:T_tau}.

\section{Charm Quark Production \label{sec:rate}}

Motivated by earlier studies~\cite{Biro:1993qt,Ko:2010zzc,Zhang:2007yoa,Song:2024hvv}, we investigate the evolution of charm quarks in the QGP by solving the differential rate equation
\begin{align}
\partial_\mu n_c^\mu  =&  R_{c\bar{c}}= R_{c\bar{c}}^{\text{gain}} - R_{c\bar{c}}^{\text{loss}},  \label{eq:rateeq} \\
R_{c\bar{c}}^{\text{gain}} =& 2\, \bar{\sigma}_{l\bar{l}\to c\bar{c}}\, \left(\frac{1}{2} n_l^{\rm eq}\right)^2 +  \bar{\sigma}_{s\bar{s}\to c\bar{c}}\, \left(n_s^{\rm eq} \right)^2 +  	\frac{1}{2}\, \bar{\sigma}_{gg\to c\bar{c}}\, \left(n_g^{\rm eq} \right)^2,
\label{eq:Rcgain}\\
R_{c\bar{c}}^{\text{loss}} =&  \Big[2 \bar{\sigma}_{c\bar{c} \to l\bar{l}}\, + \bar{\sigma}_{c\bar{c} \to s\bar{s}}\, +\frac{1}{2}\, \bar{\sigma}_{c\bar{c}\to gg}\Big] n_c^2 \label{eq:Rcloss}.
\end{align}
Equation~\eqref{eq:rateeq} describes the evolution of the charm quark number density as a result of microscopic interactions within the medium. The four-current $n_c^\mu=n_c\, u^\mu$ represents the charm number density current, where $n_c$ is the charm quark number density and $u^\mu$  is the four-velocity of the fluid. The right-hand side (RHS) of Eq.~(\ref{eq:rateeq}) corresponds to the total production rate,  $R_{c\bar{c}}$, which includes both gain and loss terms, accounting for the creation~($R_c^{\text{gain}}$) and annihilation ($R_c^{\text{loss}}$)  of the $c\bar{c}$ pairs.  In Eq.~\eqref{eq:Rcgain}, $n_i^{\rm eq}$ denotes the equilibrium number densities of the quasiparticles, with numerical prefactors ensuring the correct counting of d.o.f. for light quarks and gluons~\cite{Mykhaylova:2025gpu}. The quantities $\bar{\sigma}_{ii' \to jj'}$ are thermally averaged cross sections for the corresponding  binary scattering processes, which are discussed in detail in Sec.~\ref{sec:cross}.

In this study, we assume that only charm quarks evolve dynamically, while the quasiparticles remain in chemical equilibrium throughout the QGP lifetime. Accordingly, their equilibrium number densities are given by the standard kinetic-theory expression
\begin{align}
	n_i ^{\textrm{(eq)}}=&d_i \int \frac{{\rm d}^3p}{(2\pi)^3} f_i^{\rm{(eq)}}, \label{eq:n}
\end{align}
where the quasiparticle species $i=g,\,l,\,s$ are described by the equilibrium distributions $f_i^{\rm{eq}}$ defined in Eq.~\eqref{eq:distr}. 

Charm quarks, however, are considered out of chemical equilibrium and are therefore described by the J{\"u}ttner distribution~\cite{Biro:1993qt}
\begin{align}
	f_c[\lambda(\tau)]=\lambda(\tau)\Big[e^{E_c/T} + \lambda(\tau)\Big]^{-1} \label{eq:f_c},
\end{align}
where $\lambda(\tau)$ is a time-dependent fugacity parameter quantifying the deviation of $c$ quarks from chemical equilibrium. The evolution of the charm quark fugacity has been  studied previously in~\cite{Mykhaylova:2022toe}. This description is applied  both when treating charm quarks as impurities in a QGP with $N_f=2+1$ quark flavors as well as  when they are incorporated as thermalized quasiparticles in the $N_f=2+1+1$ scenario.

Due to their large mass, $c$ quarks can be accurately described using the Boltzmann approximation,
\begin{align}
	f_c \simeq \lambda(\tau)\, e^{-E_c/T}, \label{eq:boltz}
\end{align}
which leads to a factorized form for the number density, 
\begin{gather}
n_c[\lambda(\tau)]=\lambda(\tau) n_c^{\text{eq}}. \label{eq:factor_nc}
\end{gather}
From this relation, it is evident that as charm quarks approach equilibrium, their number  density reaches its equilibrium value, $n_c(\lambda=1)= n_c^{\text{eq}}$. Simultaneously, the production and annihilation rates in Eq.~\eqref{eq:rateeq} become equal, resulting in the conservation of the charm number density current, $\partial_\mu n_c^\mu=0$.

Using the detailed balance relation between the forward total cross sections,~$\sigma_{ii' \to  jj'}$, and their backward counterparts, $\sigma_{jj'\to ii'}$~\cite{Braun-Munzinger:2000uqj},
 \begin{align}
	\sigma_{ii' \to  jj'} =&\, \frac{d_i d_{i'}}{d_j d_{j'}} \frac{k^2_{ii'}}{k^2_{jj'}} \sigma_{jj'\to ii'}, \label{eq:en-av}\\
	k^2_{ii'(jj')}=& \,\frac{\left[s-\left(m_{i(j)}^2+m_{i'(j')}^2\right)\right] \left[s-\left(m_{i(j)}^2-m^2_{i'(j')}\right)\right]}{4s},
\end{align}
where $d_{i}$, etc.,  are the degeneracy factors defined in Sec.~\ref{sec:QPM}, and $s$ is the square of the total scattering energy, Eq.~\eqref{eq:rateeq} can be expressed in the compact form~\cite{Zhang:2007yoa,Biro:1993qt,Mykhaylova:2024xfd}
\begin{equation}
	\partial_\mu n_c^\mu[\lambda(\tau)]=R_c^{\text{gain}} \left[1-\left(\frac{n_c[\lambda(\tau)]}{n_c^{\rm eq}}\right)^2\right]\label{eq:RateEqShort}.
\end{equation}

The solution of the above equation requires knowledge of the initial charm quark fugacity,
 $	 \lambda_0	\equiv \lambda(\tau_0)$, which is determined from the SHM value for the charm yield per unit rapidity for the most central (0–10$\%$) Pb-Pb collisions at mid-rapidity, \mbox{$N_{\rm HIC}={\rm d}N_{c\bar{c}}/{\rm d}y=12.95\pm2.27$.} The uncertainty reflects the experimental error in the open charm cross section for Pb-Pb collisions~\cite{Andronic:2021erx}. 
 
 By postulating that the initial number of $c\bar{c}$ pairs in the QGP equals the value obtained from the SHM, $N_{c\bar{c}}(\tau_0)=N_{\rm HIC}$, and describing the out-of-equilibrium number density of charm quarks by Eq.~\eqref{eq:factor_nc}, $\lambda_0$ is determined as 
 \begin{gather}
\lambda_0=\frac{N_{\rm HIC}}{N_{c\bar{c}}(\tau_0)}=\frac{N_{\rm HIC}}{ n_c^{\rm eq}(\tau_0)\,V(\tau_0)},
\end{gather}
where $N_{c\bar{c}}(\tau_0)= 	 \lambda_0\, n_c^{\rm eq}(\tau_0)\, V(\tau_0)$. This yields $\lambda_0=0.0466 \pm 0.0082$ for $N_f=2+1$ and \mbox{$\lambda_0=0.0674 \pm 0.0792$} for \mbox{$N_f=2+1+1$}. The larger fugacity in the latter case reflects the increased effective quasiparticle mass of charm quarks,  $m_c(T)$, compared to the current mass $M_c$ used in the 2+1-flavor QCD (see the right panel of  Fig.~\ref{fig:GT}).

We note that the initial volume is identical in scenarios $\textbf{(a)}$ and $\textbf{(b)}$, leading to a common initial value of the charm quark fugacity at the onset of the QGP evolution.

Returning to Eq.~\eqref{eq:RateEqShort}, its LHS  explicitly depends on the number of spatial dimensions of the QGP expansion:

 $\textbf{(a)}$ For the perfect Bjorken flow (1D), we work in the fluid rest frame, where \mbox{$u^\mu=(1,0,0,0)$}, giving~\cite{Biro:1993qt,Matsui:1985eu},
\begin{eqnarray}
	\partial_\mu (n_c[\lambda(\tau)]\,u^\mu)= u^\mu \partial_\mu (n_c[\lambda(\tau)] ) + n_c[\lambda(\tau)] \partial_\mu u^\mu=\frac{\partial n_c[\lambda(\tau)]}{\partial \tau}+\frac{n_c[\lambda(\tau)]}{\tau}. \label{eq:rateLHS1D}
\end{eqnarray}

 $\textbf{(b)}$ Accounting for the transverse flow in the (2+1)D expansion and assuming a uniform density distribution in the transverse plane, the LHS of Eq.~\eqref{eq:RateEqShort} can be written as~\cite{Zhang:2007yoa,Alvarez-Ruso:2002dur,Liu:2005jb}

\begin{eqnarray}
\partial_\mu (n_c[\lambda(\tau)]\,u^\mu)= \frac{1}{\tau R^2(\tau)} \frac{\partial }{\partial \tau}\big( \tau R^2(\tau)\, n_c[\lambda(\tau)]\langle\,  u^\tau\rangle  \big), \label{eq:LHS3D}
\end{eqnarray}
where the radius $R(\tau)$ is given by Eq.~\eqref{eq:R_tau}. The flow velocity has been expressed in cylindrical coordinates, with $ \langle  u^\tau\rangle$ denoting the radial average of the flow velocity, 
\begin{gather}
	 \langle  u^\tau\rangle = \int_0^1 {\rm d}y \frac{1}{\sqrt{1-[{\rm d }R (\tau)/{\rm d}\tau]^2}y}, \label{eq:utau}
\end{gather}
where $y$ is a dummy variable. For the details of the derivation of \mbox{Eqs.~\eqref{eq:LHS3D}--\eqref{eq:utau}} we refer the reader to~\cite{Zhang:2007yoa,Alvarez-Ruso:2002dur,Liu:2005jb,Schnedermann:1993ws}.

\subsection{Total Cross Sections \label{sec:cross}}

In the QPM, heavy quarks can be created or annihilated via the following binary scatterings: $l\bar{l}\leftrightarrow c\bar{c},\ s\bar{s}\leftrightarrow c\bar{c},\ gg\leftrightarrow c\bar{c}$, described by the corresponding thermal-averaged cross sections $\bar{\sigma}_{ii'\to jj'}$.~These are obtained from the total cross  \mbox{sections ${\sigma}_{ii'\to jj'}$ via~\cite{Kadam:2015xsa,Zhang:2007yoa,Ko:1998fs,Enomoto:2023cun}}
\begin{equation}
\begin{aligned}
	\bar{\sigma}_{ii'\to jj'} =&
	\left[  8 \, \frac{m_i^2}{T^2}  \,  \frac{m_{i'}^2}{T^2}\, K_2\left( \frac{m_i}{T}\right) K_2\left( \frac{m_{i'}}{T}\right) \right]^{-1}   \\ 
&\hspace{-1.2cm}	\times\int_{\sqrt{s_0}}^\infty {\rm d}\sqrt{s}\,\left\{ \sigma_{ii'\to jj'}\   K_1 \left(\frac{\sqrt{s}}{T}\right) 
\left[\frac{s}{T^2}-\left(\frac{m_i^2}{T^2}+\frac{m_{i'}^2}{T^2}\right)^2\right]  
\left[\frac{s}{T^2}-\left(\frac{m_i^2}{T^2}-\frac{m_{i'}^2}{T^2}\right)^2\right]  \right\}, \label{sigma}
\end{aligned}
\end{equation}
where $K_{1(2)}$ are the modified Bessel functions of the first (second) kind.~The integration is performed over the total scattering energy $\sqrt{s}$, with the lower limit \sloppy 
\mbox{$ \sqrt{s_0}=\max(m_i+m_{i'},m_j+m_{j'})$}. The total cross section~${\sigma}_{ii'\to jj'}$ is computed at tree level for the on-shell quasiparticles with massive propagators. Further details can be found in~\cite{Mykhaylova:2019wci}.

Equation~(\ref{sigma}) was derived under the assumption that the scattering participants follow Boltzmann  distributions. Given the large masses of the quasiparticles and the $c$ quarks, this approximation is well justified for the temperature range considered in this study.

\begin{figure}[t]
	\centering
	\includegraphics[width=0.48\linewidth]{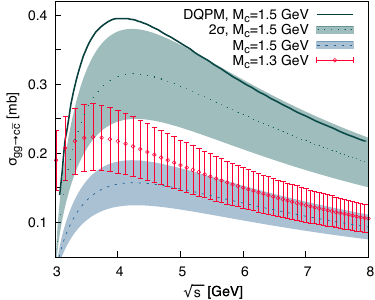}
		\includegraphics[width=0.48\linewidth]{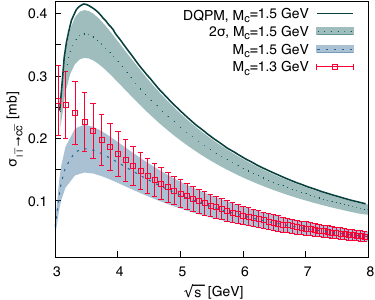}
	\caption{Total cross sections for gluon fusion, $gg\to c\bar{c}$ (left), and light quark-antiquark annihilation, $ l\bar{l}\to c\bar{c}$ (right), as functions of the collision energy $\sqrt{s}$ at $T_{\rm c}=155$ MeV for $N_f=2+1$. The~QPM results are shown for \mbox{$M_c=1.3$~GeV} (open symbols) and for $M_c=1.5$ GeV (dashed lines), as well as for the case when the cross section is multiplied by a factor of 2 (dotted lines). Alongside, the corresponding DQPM results~\cite{Song:2024hvv} are presented by solid lines. All results correspond to a thermalized QGP with $N_f=2+1$ quark flavors.}
	\label{fig:cross}
\end{figure}

In Fig.~\ref{fig:cross}, the total cross sections for $c\bar{c}$  production via gluon fusion, $\sigma_{gg\to c\bar{c}}$ (left), and light quark-antiquark annihilation, $\sigma_{ l\bar{l}\to c\bar{c}}$ (right), are shown as functions of the center-of-mass energy $\sqrt{s}$ in a QGP with $N_f=2+1$. Our results are juxtaposed to that obtained within the Dynamical Quasiparticle Model (DQPM)~\cite{Song:2024hvv}, where the cross sections of heavy quarks are doubled compared to that obtained from the tree-level Feynman diagrams~\cite{Song:2016rzw}. By adopting the same current $c$--quark mass as in the DQPM ($M_c=1.5$~GeV) and rescaling our cross sections by a factor of two, we achieve good agreement between the present QPM and the DQPM results. Since the running coupling and the corresponding quasiparticle self-energies attain very similar values in both effective frameworks, the remaining quantitative differences observed in Fig.~\ref{fig:cross} can be attributed to the finite widths of the quasiparticles incorporated in the DQPM approach.
 
The total cross sections obtained within the QPM for different values of the current \mbox{$c$--quark} mass further illustrate that, at lower energies, larger bare mass leads to a suppression of the cross sections and shifts the peak structure toward lower values of $\sqrt{s}$. At higher scattering energies, the results converge, indicating a reduced sensitivity to the choice of~$M_c$.

\subsection{Total Production Rate of Charm Quarks\label{sec:Rate-exp}}

We study the total charm quark production rate, $R_{c\bar{c}}= R_{c\bar{c}}^{\text{gain}} - R_{c\bar{c}}^{\text{loss}}$, defined in \mbox{Eqs.~\eqref{eq:Rcgain}--\eqref{eq:Rcloss},} in hot QCD matter with $N_f=2+1$, where $c$ quarks are treated as impurities with constant masses $M_c=1.3$~GeV, as well as in the $N_f=2+1+1$ case, where charm quarks are incorporated as quasiparticles with temperature-dependent masses specified by Eq.~\eqref{eq:mass}.

\begin{figure}[htb!]
	\centering
	\includegraphics[width=0.6\linewidth]{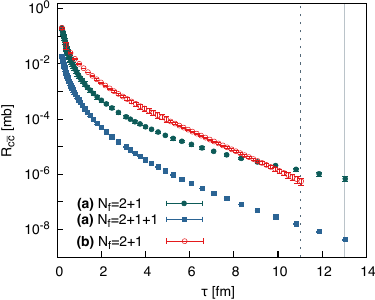}
	\caption{Total charm quark production rate  as a function of proper time, $R_{c\bar{c}}(\tau)$. Results are shown for $N_f=2+1$, where $c$ quarks are treated a heavy impurities with constant mass (circles), and for $N_f=2+1+1$ with thermalized charm quasiquarks (squares). The vertical lines indicate the times at which the QGP reaches the pseudocritical temperature, $T_{\rm c}=155$ MeV, for scenario~$\textbf{(a)}$ (perfect Bjorken flow, solid line) and $\textbf{(b)}$ (viscous medium expanding in (2+1)D, dashed line). }
	\label{fig:prodrateastau}
\end{figure}

Figure~\ref{fig:prodrateastau} shows the time dependence of the charm quark production rate in hot QCD matter for $N_f=2+1$ and $N_f=2+1+1$ cases, considering different scenarios for the QGP evolution. The $R_{c\bar{c}}(\tau)$ in the 2+1+1-flavor QGP is systematically suppressed relative to the $N_f=2+1$ case,  reflecting the larger effective mass of charm quasiparticles compared to the bare charm mass. At early times, the difference amounts to approximately one order of magnitude and increases to nearly two orders of magnitude as the system approaches the crossover region.  This behavior highlights the important role of  quasiparticle masses near~$T_{\rm c}$. While their influence is less pronounced at smaller~$\tau$ (higher $T$),  near the pseudocritical temperature, the running coupling $G(T)$ and the effective masses $m_i(T)$ become increasingly important as they encode the essential non-perturbative QCD effects. Similar trends have been reported in previous studies of the QGP transport properties, where various QPM results were found to be in agreement with the non-perturbative approaches at low~$T$, as well as with pQCD at high $T$~\cite{Mykhaylova:2019wci,Mykhaylova:2020pfk}.

For $N_f=2+1$, an overall qualitative agreement is observed between the production rates obtained in $\textbf{(a)}$ Bjorken flow, and  $\textbf{(b)}$ (2+1)D expansion based on the Bayesian analysis of HIC data~\cite{Auvinen:2020mpc}. A slight decrease of $R_{c\bar{c}}$
near the pseudocritical temperature is observed in scenario  $\textbf{(b)}$. Quantitative differences at intermediate times directly reflect the trends of the temperature profiles shown in Fig.~\ref{fig:T_tau}. 
	
The temperature dependence of the total charm production rate computed within the QPM has been extensively analyzed in~\cite{Mykhaylova:2025mht}  and compared with results from alternative effective approaches~\cite{Song:2024hvv,Levai:1997bi,Zhang:2007yoa}.  It was shown that $R_{c\bar{c}}$ computed in the underlying QPM is in agreement with available results, while exhibiting sensitivity  to the  initial conditions of the QGP evolution. Moreover, it has been reported in~\cite{Levai:1997bi,Mykhaylova:2025gpu} that the annihilation component of  $R_{c\bar{c}}$, defined in Eq.~\eqref{eq:Rcloss}, is negligible compared to the production term. This suppression originates from the large effective charm quark mass, which enters both the charm number density $n_c$ and the $c\bar{c}$ annihilation cross sections.

\section{Evolution of the Charm Quark Yield\label{sec:yield}}

We compute the time evolution of the charm quark abundance as the product of the number density defined in Eq.~\eqref{eq:factor_nc} and the system volume specified by Eq.~\eqref{eq:V},
\begin{gather}
N_{c\bar{c}}(\tau)= n_c[\lambda(\tau)] \, V(\tau). \label{eq:Nc}
\end{gather}
The charm quark fugacity $\lambda(\tau)$ is obtained by solving the rate equation~\eqref{eq:RateEqShort}.  The LHS of this equation is given by Eq.~\eqref{eq:rateLHS1D} for evolution scenario $\textbf{(a)}$, corresponding to perfect Bjorken flow, and by Eq.~\eqref{eq:LHS3D} for scenario $\textbf{(b)}$, describing a viscous QGP expanding in~(2+1)D.

\begin{figure}[t!]
	\centering
	\includegraphics[width=0.6\linewidth]{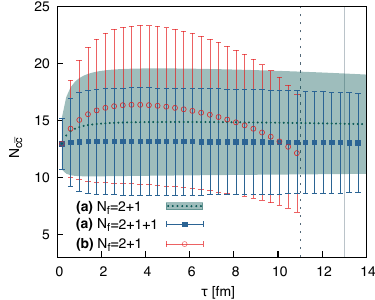}
	\caption{Total number of charm quark pairs, $N_{c\bar{c}}$,  as a function of proper time, $\tau$. In a QGP with $N_f=2+1$ quark flavors, where $c$ quarks are  treated as impurities with a constant mass $M_c=1.3$~GeV, the dotted line corresponds to perfect Bjorken flow $\textbf{(a)}$, while open circles illustrate the outcome for a viscous medium expanding  in (2+1)D $\textbf{(b)}$. The number of charm quasiquarks produced in the $N_f=2+1+1$ description is shown by squares. The~vertical lines indicate the times at which the QGP reaches the pseudocritical temperature, $T_{\rm c}=155$~MeV, for the evolution scheme $\textbf{(a)}$ (solid line) and $\textbf{(b)}$ (dashed line). All curves start from the common initial value $N_{\rm HIC}={\rm d}N_{c\bar{c}}/{\rm d}y=12.95\pm2.27$~\cite{Andronic:2021erx}.}
	\label{fig:ncastau}
\end{figure}

In Fig.~\ref{fig:ncastau}, the time evolution of the charm abundance, i.e., the total number of charm quarks in the QGP with $N_f=2+1$ quark flavors, evolving according to  scenarios $\textbf{(a)}$  and~$\textbf{(b)}$, is compared to the dynamics of $N_{c\bar{c}}$ in the QGP with $N_f=2+1+1$ obeying the scheme~$\textbf{(a)}$. We recall that the initial number of $c\bar{c}$ pairs is identical in all the cases considered, \mbox{$N_{\rm HIC}={\rm d}N_{c\bar{c}}/{\rm d}y=12.95\pm2.27$.} This corresponds to the charm yield per unit rapidity for the most central (0–10$\%$) Pb-Pb collisions at mid-rapidity, obtained within the~SHM~\cite{Andronic:2021erx}. 

$\textbf{(a)}$ For a perfect QGP propagating longitudinally, the number of charm quarks  increases slightly at very early times and then remains approximately constant until the system reaches the pseudocritical temperature. This behavior is largely independent of the treatment of heavy quarks: the evolution of $N_{c\bar{c}}$ is similar both in the $N_f=2+1$ scenario, where $c$ quarks  are treated as impurities with a fixed mass, and in the $N_f=2+1+1$ picture with dynamically dressed charm quasiparticles. Due to their effective mass, the final charm number is slightly reduced in the $N_f=2+1+1$ case,  $N_{c\bar{c}}(T_{\rm c})=13.06\pm4.43$, compared to $N_{c\bar{c}}(T_{\rm c})=14.71\pm4.42$ in the QGP with $N_f=2+1$ quark flavors.

$\textbf{(b)}$ When the hot QCD medium is described as a viscous fluid expanding in (2+1)D, the time evolution of $N_{c\bar{c}}$ exhibits a non-monotonic behavior. It initially increases, reaching a maximum of \mbox{$N_{c\bar{c}}=16.38 \pm 6.96$} at $\tau \simeq 3.5$~fm, and then decreases, resulting in a final value of \mbox{$N_{c\bar{c}}(T_{\rm c})=11.80 \pm 5.03$}. Notably, the central final value of the $c\bar{c}$ abundance is slightly smaller than the initial value, which is not observed in the QGP undergoing perfect Bjorken flow $\textbf{(a)}$. This behavior arises from the transverse acceleration parameter $a$ in Eq.~(\ref{eq:R_tau}), which affects the fireball radius~\eqref{eq:R_tau}, the system volume \eqref{eq:V}, and, consequently, the rate equation in (2+1)D \eqref{eq:LHS3D}. A similar trend, where the number of charm quarks first reaches a maximum and then decreases toward $T_{\rm c}$ was observed in~\cite{Zhang:2007yoa}, albeit for different choices of model parameters. 

Overall, the original uncertainty of $N_{\rm HIC}$, determined by the error in the open charm cross section for Pb-Pb collisions, is propagated in the QPM together with the uncertainties of the lQCD data used to fix the effective running coupling $G(T)$.  While the central final values of $N_{c\bar{c}}$ at the pseudocritical temperature differ across various QGP descriptions considered in this study, they all lie within the uncertainty of the original charm quark yield, \mbox{$N_{\rm HIC}={\rm d}N_{c\bar{c}}/{\rm d}y=12.95\pm2.27$.} This is consistent with the observation in the SHM, where the number of $c\bar{c}$ pairs remains essentially unchanged before and after the QGP phase~\cite{Andronic:2021erx}. In this context, the description of the QGP with $N_f=2+1+1$, i.e., including dynamical charm quasiparticles, yields results that are in closest agreement with the experimental data, even though the QGP evolution is simplified to a perfect Bjorken flow.

\section{Summary \label{sec:summary}}

We have studied the kinetics of charm $(c)$ quarks, including the production rate of $c\bar{c}$ pairs and the time evolution of the charm quark abundance in hot QCD medium with different quark content. In the QGP with $N_f=2+1$, charm quarks are treated as impurities of constant mass, whereas for $N_f=2+1+1$, they are fully incorporated as quasiparticles with dynamically generated masses. 
The effective quasiparticle masses are determined via the temperature-dependent running coupling, which is fixed using lattice QCD entropy density for $N_f=2+1(+1)$ \mbox{descriptions~\cite{Borsanyi:2013bia,Borsanyi:2016ksw}}. The time evolution  of the QGP is modeled using two distinct scenarios: $(\textbf{a})$ a longitudinal propagation of perfect fluid (Bjorken flow), and $(\textbf{b})$ a viscous fluid undergoing both longitudinal and transverse expansion, (2+1)D. In this case, the temperature evolution was determined from hydrodynamic simulations including a temperature-dependent specific shear viscosity constrained through a Bayesian analysis of the experimental data~\cite{Auvinen:2020mpc}.

Assuming that charm quarks can be created or annihilated through elementary two-body scattering processes, we computed the total charm quark production rate $R_{c\bar{c}}$ for both QGP compositions and evolution scenarios. In the case $\textbf{(a})$, $R_{c\bar{c}}$  for $N_f=2+1+1$  is systematically suppressed by at least one order of magnitude relative to the system with $N_f=2+1$ quark flavors due to the larger effective mass of charm quasiparticles. Moreover, for $N_f=2+1$, the charm quark production rate in the (2+1)D expansion closely follows the 1D result at early and late times but exhibits a pronounced difference at intermediate times, reflecting the distinct temperature behavior in the two scenarios.

We further studied the time evolution of the total number of $c\bar{c}$ pairs, $N_{c\bar{c}}(\tau)$, by solving the rate equation. The initial number of charm quarks corresponds to the charm yield per unit rapidity at mid-rapidity,  ${\rm d}N_{c\bar{c}}/{\rm d}y=12.95\pm2.27$, for the most central $(0-10\%)$ Pb-Pb collisions at $\sqrt{s_{NN}}=5.02$~ TeV~\cite{Andronic:2021erx}. For all QGP configurations considered, the final charm quark yield remains consistent with its initial value within the uncertainties, falling into agreement with the predictions of the Statistical Hadronization Model. In the Bjorken expansion $\textbf{(a)}$, $N_{c\bar{c}}$ increases by about 15$\%$ for $N_f=2+1$, whereas for $N_f=2+1+1$, the increase is only 1$\%$ relative to the initial number of the $c\bar{c}$ pairs. This suppression emerges from the larger effective mass of charm quasiparticles, following the dynamics observed in the production rate. In the (2+1)D expansion of the 2+1-flavor QGP, the evolution is non-monotonic, with a maximum enhancement of $\sim 25\%$ driven by the transverse acceleration. However, the excess $c\bar{c}$ pairs produced at intermediate times are subsequently lost before the system reaches the crossover region.

A more realistic study of charm quasiparticle production should incorporate both dissipative effects and transverse expansion, analogous to scenario $\textbf{(b)}$  for $N_f=2+1$. Such an analysis requires the QGP temperature evolution, which could be determined from hydrodynamic simulations employing the equation of state for $N_f=2+1+1$.

Further improvements could include next-to-leading-order scattering processes, e.g., $g\ q \leftrightarrow c\ \bar{c}\ g$ and $g\ q \leftrightarrow c\ \bar{c}\ q$~\cite{Zhang:2007yoa}, and consider finite quark chemical potential. Relaxing the assumption of chemical equilibrium for all QGP constituents would allow a more detailed study of inelastic processes and equilibration mechanisms in hot QCD medium. This, however, requires specifying initial partonic abundances, which, unlike the charm quark yield, are not directly constrained by the experimental data and are therefore intrinsically model-dependent.

\section{Acknowledgements} 
The authors thank Anton Andronic, Ilia Grishmanovskii and Pasi Huovinen for valuable comments. This work was supported by the National Science Center (NCN), Poland, under Grant No.~UMO-2021/41/N/ST2/02615. The work of K.R. and C.S. is supported by the National Science Centre (NCN), Poland, under OPUS Grant No.~2022/45/B/ST2/01527. K.R.~acknowledges the support of the Polish Ministry of Science and Higher Education. C.S.~acknowledges the support of
the World Premier International Research Center Initiative (WPI) under MEXT, Japan.

\bibliography{references}

@article{Ko:1998fs,
    author = "Ko, C. M. and Zhang, B. and Wang, X. N. and Zhang, X. F.",
    title = "{Charmonium production from hot hadronic matter}",
    eprint = "nucl-th/9808032",
    archivePrefix = "arXiv",
    doi = "10.1016/S0370-2693(98)01390-2",
    journal = "Phys. Lett. B",
    volume = "444",
    pages = "237--244",
    year = "1998"
}

@article{Sambataro:2024mkr,
    author = "Sambataro, Maria Lucia and Greco, Vincenzo and Parisi, Gabriele and Plumari, Salvatore",
    title = "{Quasi particle model vs lattice QCD thermodynamics: extension to $N_f=2+1+1$ flavors and momentum dependent quark masses}",
    eprint = "2404.17459",
    archivePrefix = "arXiv",
    primaryClass = "hep-ph",
    doi = "10.1140/epjc/s10052-024-13276-6",
    journal = "Eur. Phys. J. C",
    volume = "84",
    number = "9",
    pages = "881",
    year = "2024"
}

@article{Mykhaylova:2025gpu,
    author = "Mykhaylova, Valeriya",
    title = "{Reviewing the production rate of charm quarks in effective kinetic theory}",
    doi = "10.1016/j.jspc.2025.100031",
    journal = "J. Subatomic Part. Cosmol.",
    volume = "3",
    pages = "100031",
    year = "2025"
}

@article{Braun-Munzinger:2024ybd,
    author = "Braun-Munzinger, Peter and Redlich, Krzysztof and Sharma, Natasha and Stachel, Johanna",
    title = "{Emergence of new systematics for open charm production in high energy collisions}",
    eprint = "2408.07496",
    archivePrefix = "arXiv",
    primaryClass = "hep-ph",
    doi = "10.1007/JHEP04(2025)058",
    journal = "JHEP",
    volume = "04",
    pages = "058",
    year = "2025"
}

@article{Rapp:2008tf,
    author = "Rapp, R. and Blaschke, D. and Crochet, P.",
    title = "{Charmonium and bottomonium production in heavy-ion collisions}",
    eprint = "0807.2470",
    archivePrefix = "arXiv",
    primaryClass = "hep-ph",
    doi = "10.1016/j.ppnp.2010.07.002",
    journal = "Prog. Part. Nucl. Phys.",
    volume = "65",
    pages = "209--266",
    year = "2010"
}

@article{Beraudo:2023nlq,
    author = "Beraudo, Andrea and De Pace, Arturo and Pablos, Daniel and Prino, Francesco and Monteno, Marco and Nardi, Marzia",
    title = "{Heavy-flavor transport and hadronization in pp collisions}",
    eprint = "2306.02152",
    archivePrefix = "arXiv",
    primaryClass = "hep-ph",
    doi = "10.1103/PhysRevD.109.L011501",
    journal = "Phys. Rev. D",
    volume = "109",
    number = "1",
    pages = "L011501",
    year = "2024"
}

@article{Dang:2023tmb,
    author = "Dang, Yichao and Xing, Wen-Jing and Cao, Shanshan and Qin, Guang-You",
    title = "{Reexamining charm versus bottom quark energy loss inside a color-deconfined medium}",
    eprint = "2307.14808",
    archivePrefix = "arXiv",
    primaryClass = "nucl-th",
    doi = "10.1103/PhysRevC.109.064901",
    journal = "Phys. Rev. C",
    volume = "109",
    number = "6",
    pages = "064901",
    year = "2024"
}

@article{Xing:2023ciw,
    author = "Xing, Wen-Jing and Cao, Shanshan and Qin, Guang-You",
    title = "{Flavor hierarchy of parton energy loss in quark-gluon plasma from a Bayesian analysis}",
    eprint = "2303.12485",
    archivePrefix = "arXiv",
    primaryClass = "hep-ph",
    doi = "10.1016/j.physletb.2024.138523",
    journal = "Phys. Lett. B",
    volume = "850",
    pages = "138523",
    year = "2024"
}

@article{ALICE:2023jgm,
    author = "Acharya, Shreyasi and others",
    collaboration = "ALICE",
    title = "{Measurement of the fraction of jet longitudinal momentum carried by {\ensuremath{\Lambda}}c+ baryons in pp collisions}",
    eprint = "2301.13798",
    archivePrefix = "arXiv",
    primaryClass = "nucl-ex",
    reportNumber = "CERN-EP-2023-005",
    doi = "10.1103/PhysRevD.109.072005",
    journal = "Phys. Rev. D",
    volume = "109",
    number = "7",
    pages = "072005",
    year = "2024"
}

@article{Beraudo:2025nvq,
    author = "Beraudo, Andrea and Du Plessis, Jean F. and Pablos, Daniel and Rajagopal, Krishna",
    title = "{Heavy Quark Energy Loss in the Hybrid Model}",
    eprint = "2510.24847",
    archivePrefix = "arXiv",
    primaryClass = "hep-ph",
    reportNumber = "MIT-CTP/5950",
    month = "10",
    year = "2025"
}

@article{Graf:2018lok,
    author = "Graf, Thorben and Steinheimer, Jan and Bleicher, Marcus and Herold, Christoph",
    title = "{Testing charm quark equilibration in ultrahigh-energy heavy ion collisions with fluctuations}",
    eprint = "1802.07908",
    archivePrefix = "arXiv",
    primaryClass = "hep-ph",
    doi = "10.1103/PhysRevC.97.034906",
    journal = "Phys. Rev. C",
    volume = "97",
    number = "3",
    pages = "034906",
    year = "2018"
}

@article{Uphoff:2010sh,
    author = "Uphoff, Jan and Fochler, Oliver and Xu, Zhe and Greiner, Carsten",
    title = "{Heavy quark production at RHIC and LHC within a partonic transport model}",
    eprint = "1003.4200",
    archivePrefix = "arXiv",
    primaryClass = "hep-ph",
    doi = "10.1103/PhysRevC.82.044906",
    journal = "Phys. Rev. C",
    volume = "82",
    pages = "044906",
    year = "2010"
}

@article{Apolinario:2022vzg,
    author = "Apolin{\'a}rio, Liliana and Lee, Yen-Jie and Winn, Michael",
    title = "{Heavy quarks and jets as probes of the QGP}",
    eprint = "2203.16352",
    archivePrefix = "arXiv",
    primaryClass = "hep-ph",
    doi = "10.1016/j.ppnp.2022.103990",
    journal = "Prog. Part. Nucl. Phys.",
    volume = "127",
    pages = "103990",
    year = "2022"
}

@article{Gross:2022hyw,
    author = "Gross, Franz and others",
    title = "{50 Years of Quantum Chromodynamics}",
    eprint = "2212.11107",
    archivePrefix = "arXiv",
    primaryClass = "hep-ph",
    doi = "10.1140/epjc/s10052-023-11949-2",
    journal = "Eur. Phys. J. C",
    volume = "83",
    pages = "1125",
    year = "2023"
}

@article{Berwein:2024ztx,
    author = "Berwein, Matthias and Brambilla, Nora and Mohapatra, Abhishek and Vairo, Antonio",
    title = "{Hybrids, tetraquarks, pentaquarks, doubly heavy baryons, and quarkonia in Born-Oppenheimer effective theory}",
    eprint = "2408.04719",
    archivePrefix = "arXiv",
    primaryClass = "hep-ph",
    reportNumber = "TUM-EFT 185/23",
    doi = "10.1103/PhysRevD.110.094040",
    journal = "Phys. Rev. D",
    volume = "110",
    number = "9",
    pages = "094040",
    year = "2024"
}

@article{Andronic:2025jbp,
    author = "Andronic, Anton and Arnaldi, Roberta",
    title = "{Quarkonia and Deconfined Quark{\textendash}Gluon Matter in Heavy-Ion Collisions}",
    eprint = "2501.08290",
    archivePrefix = "arXiv",
    primaryClass = "nucl-ex",
    doi = "10.1146/annurev-nucl-121423-101041",
    journal = "Ann. Rev. Nucl. Part. Sci.",
    volume = "75",
    number = "1",
    pages = "351--375",
    year = "2025"
}

@article{HotQCD:2025fbd,
    author = "Bollweg, Dennis and Dasilva Gol{\'a}n, Jorge Luis and Kaczmarek, Olaf and Larsen, Rasmus Norman and Moore, Guy D. and Mukherjee, Swagato and Petreczky, Peter and Shu, Hai-Tao and Stendebach, Simon and Weber, Johannes Heinrich",
    collaboration = "HotQCD",
    title = "{Temperature dependence of heavy quark diffusion from (2+1)-flavor lattice QCD}",
    eprint = "2506.11958",
    archivePrefix = "arXiv",
    primaryClass = "hep-lat",
    doi = "10.1007/JHEP09(2025)180",
    journal = "JHEP",
    volume = "09",
    pages = "180",
    year = "2025"
}

@article{Zhao:2023nrz,
    author = "Zhao, Jiaxing and others",
    title = "{Hadronization of heavy quarks}",
    eprint = "2311.10621",
    archivePrefix = "arXiv",
    primaryClass = "hep-ph",
    doi = "10.1103/PhysRevC.109.054912",
    journal = "Phys. Rev. C",
    volume = "109",
    number = "5",
    pages = "054912",
    year = "2024"
}

@article{Altenkort:2023oms,
    author = "Altenkort, Luis and Kaczmarek, Olaf and Larsen, Rasmus and Mukherjee, Swagato and Petreczky, Peter and Shu, Hai-Tao and Stendebach, Simon",
    collaboration = "HotQCD",
    title = "{Heavy Quark Diffusion from 2+1 Flavor Lattice QCD with 320~MeV Pion Mass}",
    eprint = "2302.08501",
    archivePrefix = "arXiv",
    primaryClass = "hep-lat",
    doi = "10.1103/PhysRevLett.130.231902",
    journal = "Phys. Rev. Lett.",
    volume = "130",
    number = "23",
    pages = "231902",
    year = "2023"
}

@article{Krishna:2025bll,
    author = "Krishna, Tharun and Rapp, Ralf and Fu, Yu and Bass, Steffen A. and Ke, Weiyao",
    title = "{Nonperturbative heavy-flavor transport approach for hot QCD matter}",
    eprint = "2509.13881",
    archivePrefix = "arXiv",
    primaryClass = "nucl-th",
    doi = "10.1016/j.physletb.2025.139999",
    journal = "Phys. Lett. B",
    volume = "871",
    pages = "139999",
    year = "2025"
}

@article{He:2022ywp,
    author = "He, Min and van Hees, Hendrik and Rapp, Ralf",
    title = "{Heavy-quark diffusion in the quark{\textendash}gluon plasma}",
    eprint = "2204.09299",
    archivePrefix = "arXiv",
    primaryClass = "hep-ph",
    doi = "10.1016/j.ppnp.2023.104020",
    journal = "Prog. Part. Nucl. Phys.",
    volume = "130",
    pages = "104020",
    year = "2023"
}

@article{Dong:2019byy,
    author = "Dong, Xin and Lee, Yen-Jie and Rapp, Ralf",
    title = "{Open Heavy-Flavor Production in Heavy-Ion Collisions}",
    eprint = "1903.07709",
    archivePrefix = "arXiv",
    primaryClass = "nucl-ex",
    doi = "10.1146/annurev-nucl-101918-023806",
    journal = "Ann. Rev. Nucl. Part. Sci.",
    volume = "69",
    pages = "417--445",
    year = "2019"
}

@article{Bierlich:2023ewv,
    author = "Bierlich, Christian and Wilkinson, Jeremy and Sun, Jiayin and Manca, Giulia and Granier de Cassagnac, Raphael and Otwinowski, Jacek",
    title = "{Open charm production cross section from combined LHC experiments in pp collisions at $\sqrt{s} = 5.02$~TeV}",
    eprint = "2311.11426",
    archivePrefix = "arXiv",
    primaryClass = "hep-ph",
    doi = "10.1140/epjp/s13360-024-05355-0",
    journal = "Eur. Phys. J. Plus",
    volume = "139",
    number = "7",
    pages = "593",
    year = "2024"
}

@article{LHCb:2023cwu,
    author = "Aaij, Roel and others",
    collaboration = "LHCb",
    title = "{Measurement of {\ensuremath{\Xi}}c+ production in pPb collisions at sNN=8.16 TeV at LHCb}",
    eprint = "2305.06711",
    archivePrefix = "arXiv",
    primaryClass = "hep-ex",
    reportNumber = "LHCb-PAPER-2022-041, CERN-EP-2023-057",
    doi = "10.1103/PhysRevC.109.044901",
    journal = "Phys. Rev. C",
    volume = "109",
    number = "4",
    pages = "044901",
    year = "2024"
}

@article{LHCb:2022dmh,
    author = "Aaij, R. and others",
    collaboration = "LHCb",
    title = "{Measurement of the Prompt D0 Nuclear Modification Factor in p-Pb Collisions at sNN=8.16{\,}{\,}TeV}",
    eprint = "2205.03936",
    archivePrefix = "arXiv",
    primaryClass = "nucl-ex",
    reportNumber = "LHCb-PAPER-2022-007, CERN-EP-2022-082",
    doi = "10.1103/PhysRevLett.131.102301",
    journal = "Phys. Rev. Lett.",
    volume = "131",
    number = "10",
    pages = "102301",
    year = "2023"
}

@article{Rajagopal:2025ukd,
    author = "Rajagopal, Krishna and Scheihing-Hitschfeld, Bruno and Wiedemann, Urs Achim",
    title = "{Dynamics of heavy quarks in strongly coupled $ \mathcal{N} $ = 4 SYM plasma}",
    eprint = "2501.06289",
    archivePrefix = "arXiv",
    primaryClass = "hep-ph",
    reportNumber = "CERN-TH-2025-008, MIT-CTP/5822",
    doi = "10.1007/JHEP07(2025)013",
    journal = "JHEP",
    volume = "07",
    pages = "013",
    year = "2025"
}

@article{Altenkort:2023eav,
    author = "Altenkort, Luis and de la Cruz, David and Kaczmarek, Olaf and Larsen, Rasmus and Moore, Guy D. and Mukherjee, Swagato and Petreczky, Peter and Shu, Hai-Tao and Stendebach, Simon",
    collaboration = "HotQCD",
    title = "{Quark Mass Dependence of Heavy Quark Diffusion Coefficient from Lattice QCD}",
    eprint = "2311.01525",
    archivePrefix = "arXiv",
    primaryClass = "hep-lat",
    doi = "10.1103/PhysRevLett.132.051902",
    journal = "Phys. Rev. Lett.",
    volume = "132",
    number = "5",
    pages = "051902",
    year = "2024"
}

@article{Rapp:2008qc,
    author = "Rapp, Ralf and van Hees, Hendrik",
    title = "{Heavy Quark Diffusion as a Probe of the Quark-Gluon Plasma}",
    eprint = "0803.0901",
    archivePrefix = "arXiv",
    primaryClass = "hep-ph",
    month = "3",
    year = "2008"
}

@article{Singh:2025dfx,
    author = "Singh, Sunny Kumar and Bhadury, Samapan and Ghosh, Ritesh and Kurian, Manu",
    title = "{Quarkonium in a QCD medium with momentum-dependent relaxation time}",
    eprint = "2508.09108",
    archivePrefix = "arXiv",
    primaryClass = "hep-ph",
    doi = "10.1103/rb44-j8p3",
    journal = "Phys. Rev. D",
    volume = "112",
    number = "9",
    pages = "094056",
    year = "2025"
}

@article{Zhao:2023ucp,
    author = "Zhao, Jiaxing and Aichelin, Joerg and Gossiaux, Pol Bernard and Werner, Klaus",
    title = "{Heavy flavor as a probe of hot QCD matter produced in proton-proton collisions}",
    eprint = "2310.08684",
    archivePrefix = "arXiv",
    primaryClass = "hep-ph",
    doi = "10.1103/PhysRevD.109.054011",
    journal = "Phys. Rev. D",
    volume = "109",
    number = "5",
    pages = "054011",
    year = "2024"
}

@article{CMS:2018loe,
    author = "Sirunyan, A. M. and others",
    collaboration = "CMS",
    title = "{Elliptic flow of charm and strange hadrons in high-multiplicity pPb collisions at $\sqrt{s_{_\mathrm{NN}}} =$ 8.16 TeV}",
    eprint = "1804.09767",
    archivePrefix = "arXiv",
    primaryClass = "hep-ex",
    reportNumber = "CMS-HIN-17-003, CERN-EP-2018-076",
    doi = "10.1103/PhysRevLett.121.082301",
    journal = "Phys. Rev. Lett.",
    volume = "121",
    number = "8",
    pages = "082301",
    year = "2018"
}

@article{Peng:2024zvf,
    author = "Peng, Jiazhen and Yu, Kewei and Li, Shuang and Xiong, Wei and Sun, Fei and Xie, Wei",
    title = "{Unraveling the collisional energy loss of a heavy quark in a quark-gluon plasma}",
    eprint = "2401.10644",
    archivePrefix = "arXiv",
    primaryClass = "hep-ph",
    doi = "10.1103/PhysRevD.109.096028",
    journal = "Phys. Rev. D",
    volume = "109",
    number = "9",
    pages = "096028",
    year = "2024"
}

@article{Moore:2004tg,
    author = "Moore, Guy D. and Teaney, Derek",
    title = "{How much do heavy quarks thermalize in a heavy ion collision?}",
    eprint = "hep-ph/0412346",
    archivePrefix = "arXiv",
    doi = "10.1103/PhysRevC.71.064904",
    journal = "Phys. Rev. C",
    volume = "71",
    pages = "064904",
    year = "2005"
}

@article{Yao:2020xzw,
    author = {Yao, Xiaojun and Ke, Weiyao and Xu, Yingru and Bass, Steffen A. and M{\"u}ller, Berndt},
    title = "{Coupled Boltzmann Transport Equations of Heavy Quarks and Quarkonia in Quark-Gluon Plasma}",
    eprint = "2004.06746",
    archivePrefix = "arXiv",
    primaryClass = "hep-ph",
    reportNumber = "MIT-CTP/5192",
    doi = "10.1007/JHEP01(2021)046",
    journal = "JHEP",
    volume = "01",
    pages = "046",
    year = "2021"
}

@inbook{Rapp:2009my,
author = {Ralf Rapp and Hendrik van Hees},
title = {HEAVY QUARKS IN THE QUARK-GLUON PLASMA},
booktitle = {Quark-Gluon Plasma 4},
chapter = {},
pages = {111-206},
eprint = "0903.1096",
    archivePrefix = "arXiv",
    primaryClass = "hep-ph",
    doi = "10.1142/9789814293297_0003"}

@article{Scardina:2017ipo,
    author = "Scardina, Francesco and Das, Santosh K. and Minissale, Vincenzo and Plumari, Salvatore and Greco, Vincenzo",
    title = "{Estimating the charm quark diffusion coefficient and thermalization time from D meson spectra at energies available at the BNL Relativistic Heavy Ion Collider and the CERN Large Hadron Collider}",
    eprint = "1707.05452",
    archivePrefix = "arXiv",
    primaryClass = "nucl-th",
    doi = "10.1103/PhysRevC.96.044905",
    journal = "Phys. Rev. C",
    volume = "96",
    number = "4",
    pages = "044905",
    year = "2017"
}

@article{Zhao:2011cv,
    author = "Zhao, Xingbo and Rapp, Ralf",
    title = "{Medium Modifications and Production of Charmonia at LHC}",
    eprint = "1102.2194",
    archivePrefix = "arXiv",
    primaryClass = "hep-ph",
    doi = "10.1016/j.nuclphysa.2011.05.001",
    journal = "Nucl. Phys. A",
    volume = "859",
    pages = "114--125",
    year = "2011"
}

@article{Ding:2012sp,
    author = "Ding, H. T. and Francis, A. and Kaczmarek, O. and Karsch, F. and Satz, H. and Soeldner, W.",
    title = "{Charmonium properties in hot quenched lattice QCD}",
    eprint = "1204.4945",
    archivePrefix = "arXiv",
    primaryClass = "hep-lat",
    reportNumber = "BI-TP-2012-13",
    doi = "10.1103/PhysRevD.86.014509",
    journal = "Phys. Rev. D",
    volume = "86",
    pages = "014509",
    year = "2012"
}

@article{Capellino:2023cxe,
    author = "Capellino, F. and Dubla, A. and Floerchinger, S. and Grossi, E. and Kirchner, A. and Masciocchi, S.",
    title = "{Fluid dynamics of charm quarks in the quark-gluon plasma}",
    eprint = "2307.14449",
    archivePrefix = "arXiv",
    primaryClass = "hep-ph",
    doi = "10.1103/PhysRevD.108.116011",
    journal = "Phys. Rev. D",
    volume = "108",
    number = "11",
    pages = "116011",
    year = "2023"
}

@article{Sambataro:2025obe,
    author = "Sambataro, Maria Lucia and Minissale, Vincenzo and Plumari, Salvatore and Greco, Vincenzo",
    title = "{Assessing lattice QCD charm space diffusion coefficient and thermalization time by mean of D meson observables at LHC}",
    eprint = "2508.01024",
    archivePrefix = "arXiv",
    primaryClass = "hep-ph",
    doi = "10.1016/j.physletb.2025.140049",
    journal = "Phys. Lett. B",
    volume = "872",
    pages = "140049",
    year = "2026"
}

@article{Bluhm:2004xn,
    author = "Bluhm, M. and Kampfer, Burkhard and Soff, G.",
    title = "{The QCD equation of state near T(c) within a quasi-particle model}",
    eprint = "hep-ph/0411106",
    archivePrefix = "arXiv",
    doi = "10.1016/j.physletb.2005.05.083",
    journal = "Phys. Lett. B",
    volume = "620",
    pages = "131--136",
    year = "2005"
}

@article{Bluhm:2007nu,
    author = "Bluhm, M. and Kampfer, Burkhard and Schulze, R. and Seipt, D. and Heinz, U.",
    title = "{A family of equations of state based on lattice QCD: Impact on flow in ultrarelativistic heavy-ion collisions}",
    eprint = "0705.0397",
    archivePrefix = "arXiv",
    primaryClass = "hep-ph",
    doi = "10.1103/PhysRevC.76.034901",
    journal = "Phys. Rev. C",
    volume = "76",
    pages = "034901",
    year = "2007"
}

@article{Liu:2005jb,
    author = "Liu, W. and Ko, C. M. and Chen, L. W.",
    title = "{Eta absorption by mesons}",
    eprint = "nucl-th/0505075",
    archivePrefix = "arXiv",
    doi = "10.1016/j.nuclphysa.2005.11.008",
    journal = "Nucl. Phys. A",
    volume = "765",
    pages = "401--425",
    year = "2006"
}

@article{Song:2016rzw,
    author = "Song, Taesoo and Berrehrah, Hamza and Torres-Rincon, Juan M. and Tolos, Laura and Cabrera, Daniel and Cassing, Wolfgang and Bratkovskaya, Elena",
    title = "{Single electrons from heavy-flavor mesons in relativistic heavy-ion collisions}",
    eprint = "1605.07887",
    archivePrefix = "arXiv",
    primaryClass = "nucl-th",
    doi = "10.1103/PhysRevC.96.014905",
    journal = "Phys. Rev. C",
    volume = "96",
    number = "1",
    pages = "014905",
    year = "2017"
}

@article{Borsanyi:2016ksw,
    author = "Borsanyi, Sz. and others",
    title = "{Calculation of the axion mass based on high-temperature lattice quantum chromodynamics}",
    eprint = "1606.07494",
    archivePrefix = "arXiv",
    primaryClass = "hep-lat",
    reportNumber = "DESY-16-105",
    doi = "10.1038/nature20115",
    journal = "Nature",
    volume = "539",
    number = "7627",
    pages = "69--71",
    year = "2016"
}

@article{Mykhaylova:2021cep,
    author = "Mykhaylova, V.",
    title = "{Interviewing the Weak with Strong Coupling Regimes via the Bulk to Shear Viscosity Ratio in QCD}",
    doi = "10.5506/aphyspolbsupp.14.271",
    journal = "Acta Phys. Polon. Supp.",
    volume = "14",
    number = "2",
    year = "2021"
}

@article{Mykhaylova:2024xfd,
    author = "Mykhaylova, V.",
    title = "{Production Rate of Charm Quarks in the Quasiparticle Approach}",
    doi = "10.5506/APhysPolBSupp.17.6-A10",
    journal = "Acta Phys. Polon. Supp.",
    volume = "17",
    number = "6",
    pages = "6--A10",
    year = "2024"
}

@article{Song:2024hvv,
    author = "Song, Taesoo and Grishmanovskii, Ilia and Soloveva, Olga and Bratkovskaya, Elena",
    title = "{Thermal production of charm quarks in relativistic heavy-ion collisions}",
    eprint = "2404.00425",
    archivePrefix = "arXiv",
    primaryClass = "nucl-th",
    doi = "10.1103/PhysRevC.110.034906",
    journal = "Phys. Rev. C",
    volume = "110",
    number = "3",
    pages = "034906",
    year = "2024"
}

@article{Mykhaylova:2025mht,
    author = "Mykhaylova, Valeriya",
    title = "{Charm Quark Evolution in the Quark{\textendash}Gluon Plasma with Various Quark Contents}",
    doi = "10.3390/physics7030039",
    journal = "MDPI Physics",
    volume = "7",
    number = "3",
    pages = "39",
    year = "2025"
}

@article{Kadam:2015xsa,
    author = "Kadam, Guru Prakash and Mishra, Hiranmaya",
    title = "{Dissipative properties of hot and dense hadronic matter in an excluded-volume hadron resonance gas model}",
    eprint = "1506.04613",
    archivePrefix = "arXiv",
    primaryClass = "hep-ph",
    doi = "10.1103/PhysRevC.92.035203",
    journal = "Phys. Rev. C",
    volume = "92",
    number = "3",
    pages = "035203",
    year = "2015"
}

@article{Biro:1993qt,
    author = "Biro, T. S. and van Doorn, E. and Muller, Berndt and Thoma, M. H. and Wang, X. N.",
    title = "{Parton equilibration in relativistic heavy ion collisions}",
    eprint = "nucl-th/9303004",
    archivePrefix = "arXiv",
    reportNumber = "DUKE-TH-93-46",
    doi = "10.1103/PhysRevC.48.1275",
    journal = "Phys. Rev. C",
    volume = "48",
    pages = "1275--1284",
    year = "1993"
}

@article{Bjorken:1982qr,
  title = {Highly relativistic nucleus-nucleus collisions: The central rapidity region},
  author = {Bjorken, J. D.},
  journal = {Phys. Rev. D},
  volume = {27},
  issue = {1},
  pages = {140--151},
  numpages = {0},
  year = {1983},
  month = {Jan},
  publisher = {American Physical Society},
  doi = {10.1103/PhysRevD.27.140},
  url = {https://link.aps.org/doi/10.1103/PhysRevD.27.140}
}

@article{Matsui:1985eu,
    author = "Matsui, T. and Svetitsky, B. and McLerran, Larry D.",
    title = "{Strangeness Production in Ultrarelativistic Heavy Ion Collisions. 1. Chemical Kinetics in the Quark - Gluon Plasma}",
    reportNumber = "MIT-CTP-1320",
    doi = "10.1103/PhysRevD.37.844",
    journal = "Phys. Rev. D",
    volume = "34",
    pages = "783",
    year = "1986",
    note = "[Erratum: Phys.Rev.D 37, 844 (1988)]"
}

@article{Zhang:2007yoa,
    author = "Zhang, B.-W. and Ko, C.-M. and Liu, W.",
    title = "{Thermal charm production in a quark-gluon plasma in Pb-Pb collisions at s**(1/2)(NN) = 5.5-TeV}",
    eprint = "0709.1684",
    archivePrefix = "arXiv",
    primaryClass = "nucl-th",
    doi = "10.1103/PhysRevC.77.024901",
    journal = "Phys. Rev. C",
    volume = "77",
    pages = "024901",
    year = "2008"
}

@article{Auvinen:2020mpc,
    author = "Auvinen, Jussi and Eskola, Kari J. and Huovinen, Pasi and Niemi, Harri and Paatelainen, Risto and Petreczky, Peter",
    title = "{Temperature dependence of $\eta/s$ of strongly interacting matter: Effects of the equation of state and the parametric form of $(\eta/s)(T)$}",
    eprint = "2006.12499",
    archivePrefix = "arXiv",
    primaryClass = "nucl-th",
    reportNumber = "CERN-TH-2020-099",
    doi = "10.1103/PhysRevC.102.044911",
    journal = "Phys. Rev. C",
    volume = "102",
    number = "4",
    pages = "044911",
    year = "2020"
}

@article{Mykhaylova:2019wci,
    author = "Mykhaylova, V. and Bluhm, M. and Redlich, K. and Sasaki, C.",
    title = "{Quark-flavor dependence of the shear viscosity in a quasiparticle model}",
    eprint = "1906.01697",
    archivePrefix = "arXiv",
    primaryClass = "hep-ph",
    doi = "10.1103/PhysRevD.100.034002",
    journal = "Phys. Rev. D",
    volume = "100",
    number = "3",
    pages = "034002",
    year = "2019"
}

@article{Mykhaylova:2020pfk,
    author = "Mykhaylova, V. and Sasaki, C.",
    title = "{Impact of quark quasiparticles on transport coefficients in hot QCD}",
    eprint = "2007.06846",
    archivePrefix = "arXiv",
    primaryClass = "hep-ph",
    doi = "10.1103/PhysRevD.103.014007",
    journal = "Phys. Rev. D",
    volume = "103",
    number = "1",
    pages = "014007",
    year = "2021"
}

@article{Schnedermann:1993ws,
    author = "Schnedermann, Ekkard and Sollfrank, Josef and Heinz, Ulrich W.",
    title = "{Thermal phenomenology of hadrons from 200-A/GeV S+S collisions}",
    eprint = "nucl-th/9307020",
    archivePrefix = "arXiv",
    reportNumber = "TPR-93-16",
    doi = "10.1103/PhysRevC.48.2462",
    journal = "Phys. Rev. C",
    volume = "48",
    pages = "2462--2475",
    year = "1993"
}

@article{Andronic:2021erx,
    author = {Andronic, Anton and Braun-Munzinger, Peter and K{\"o}hler, Markus K. and Mazeliauskas, Aleksas and Redlich, Krzysztof and Stachel, Johanna and Vislavicius, Vytautas},
    title = "{The multiple-charm hierarchy in the statistical hadronization model}",
    eprint = "2104.12754",
    archivePrefix = "arXiv",
    primaryClass = "hep-ph",
    doi = "10.1007/JHEP07(2021)035",
    journal = "JHEP",
    volume = "07",
    pages = "035",
    year = "2021"
}

@article{Ko:2010zzc,
    author = "Ko, Che Ming and Lee, Su Houng and Liu, Wei and Oh, Yongseok and Yasui, Shigehiro and Zhang, Ben-Wei",
    editor = "Blaschke, David and Redlich, Krzysztof and Turko, Ludwik and Zablocki, Daniel",
    title = "{Charms in relativistic heavy-ion collisions}",
    journal = "Acta Phys. Polon. Supp.",
    volume = "3",
    pages = "601--610",
    year = "2010"
}

@article{Sheikh:2021opp,
    author = "Sheikh, Ashik Ikbal",
    title = "{Effect of thermal gluon absorption and medium fluctuations on heavy flavour nuclear modification factor at RHIC and LHC energies}",
    eprint = "2206.14243",
    archivePrefix = "arXiv",
    primaryClass = "hep-ph",
    doi = "10.1140/epja/s10050-021-00636-z",
    journal = "Eur. Phys. J. A",
    volume = "57",
    number = "12",
    pages = "323",
    year = "2021"
}

@article{Kumar:2014kfa,
    author = "Kumar, Vineet and Shukla, Prashant and Vogt, Ramona",
    title = "{Quarkonia suppression in PbPb collisions at $\sqrt{s_{NN}}$ = 2.76 TeV}",
    eprint = "1410.3299",
    archivePrefix = "arXiv",
    primaryClass = "hep-ph",
    doi = "10.1103/PhysRevC.92.024908",
    journal = "Phys. Rev. C",
    volume = "92",
    number = "2",
    pages = "024908",
    year = "2015"
}

@article{Alvarez-Ruso:2002dur,
    author = "Alvarez-Ruso, L. and Koch, V.",
    title = "{Phi meson propagation in a hot hadronic gas}",
    eprint = "nucl-th/0201011",
    archivePrefix = "arXiv",
    doi = "10.1103/PhysRevC.65.054901",
    journal = "Phys. Rev. C",
    volume = "65",
    pages = "054901",
    year = "2002"
}

@article{Bluhm:2006yh,
    author = "Bluhm, Marcus and Kampfer, Burkhard and Schulze, Robert and Seipt, Daniel",
    editor = "Antinori, F. and Bass, S. and De Falco, A. and Kuhn, C. and Nardi, M. and Peitzmann, T. and Ullrich, T. and Velkovska, J. and Wiedemann, U. A.",
    title = "{Quasi-Particle Description of Strongly Interacting Matter: Towards a Foundation}",
    eprint = "hep-ph/0608053",
    archivePrefix = "arXiv",
    doi = "10.1140/epjc/s10052-006-0056-y",
    journal = "Eur. Phys. J. C",
    volume = "49",
    pages = "205--211",
    year = "2007"
}

@article{Pisarski:1989wb,
    author = "Pisarski, R. D.",
    editor = "Baym, G. A. and Braun-Munzinger, P. and Nagamiya, S.",
    title = "{Renormalized Fermion Propagator in Hot Gauge Theories}",
    doi = "10.1016/0375-9474(89)90620-9",
    journal = "Nucl. Phys. A",
    volume = "498",
    pages = "423C--428C",
    year = "1989"
}

@article{Borsanyi:2013bia,
    author = "Borsanyi, S. and Fodor, Z. and Hoelbling, C. and Katz, S. D. and Krieg, S. and Szabo, K. K.",
    title = "{Full result for the QCD equation of state with 2+1 flavors}",
    eprint = "1309.5258",
    archivePrefix = "arXiv",
    primaryClass = "hep-lat",
    doi = "10.1016/j.physletb.2014.01.007",
    journal = "Phys. Lett. B",
    volume = "730",
    pages = "99--104",
    year = "2014"
}

@article{Levai:1997bi,
    author = "Levai, P. and Vogt, R.",
    title = "{Thermal charm production by massive gluons and quarks}",
    eprint = "hep-ph/9704360",
    archivePrefix = "arXiv",
    reportNumber = "LBL-39984, LBNL-39984",
    doi = "10.1103/PhysRevC.56.2707",
    journal = "Phys. Rev. C",
    volume = "56",
    pages = "2707--2717",
    year = "1997"
}

@article{Mykhaylova:2022toe,
    author = "Mykhaylova, V.",
    title = "{Charm quark fugacity in hot QCD}",
    doi = "10.1051/epjconf/202227405006",
    journal = "EPJ Web Conf.",
    volume = "274",
    pages = "05006",
    year = "2022"
}

@Article{Enomoto:2023cun,
AUTHOR = {Enomoto, Seishi and Su, Yu-Hang and Zheng, Man-Zhu and Zhang, Hong-Hao},
TITLE = {Boltzmann Equation and Its Cosmological Applications},
JOURNAL = {Symmetry},
VOLUME = {17},
YEAR = {2025},
NUMBER = {6},
ARTICLE-NUMBER = {921},
URL = {https://www.mdpi.com/2073-8994/17/6/921},
ISSN = {2073-8994},
DOI = {10.3390/sym17060921}
}

@article{Braun-Munzinger:2000uqj,
    author = "Braun-Munzinger, Peter and Redlich, Krzysztof",
    title = "{Charmonium production from the secondary collisions at LHC energy}",
    eprint = "hep-ph/0001008",
    archivePrefix = "arXiv",
    doi = "10.1007/s100520000356",
    journal = "Eur. Phys. J. C",
    volume = "16",
    pages = "519--525",
    year = "2000"
}
\end{document}